# Sparse Signal Recovery from Nonadaptive Linear Measurements

Ankit Kundu
IBM India Pvt. Ltd ,
9th Floor DLF IT Park, Calcutta 700156, India
ankitkundu931@yahoo.in

Pradosh K. Roy
Asia Pacific Institute of Information Technology,
Panipat, Haryana 132103 , India
roypkin@yahoo.com

*Abstract :  The theory of Compressed Sensing  , the emerging sampling paradigm 'that goes against the common wisdom' , asserts that 'one can recover signals in $\mathbb{R}^n$  from far fewer samples or measurements ,  if the signal has  a sparse representation in some orthonormal basis,  from $m \approx O(k \log n)$, $k \ll n$  nonadaptive measurements . The accuracy of the recovered signal is as good as that attainable with direct knowledge of the k most important coefficients and its locations. Moreover, a good approximation to those  important coefficients is extracted from the measurements by solving a $\ell_1$- minimization  problem - Basis Pursuit. ' The nonadaptive  measurements have the character of "random" linear combinations of the basis/frame elements'.*

*The  theory has  implications which are far reaching  and immediately leads to a number of  applications in Data Compression ,Channel Coding and   Data Acquisition. 'The last of these applications suggest that CS could have an enormous impact in areas where conventional hardware design has significant  limitations' ,  leading  to  ' efficient and revolutionary methods of data acquisition and storage  in future'.*

*The paper reviews fundamental  mathematical ideas pertaining to compressed sensing viz. sparsity, incoherence , reduced isometry property and basis pursuit ,  exemplified by  the sparse recovery of a speech signal and  convergence of  the $\ell_1$- minimization algorithm.*

## I. INTRODUCTION

A conventional approach in digital signal acquisition and processing is to assume that the signal is bandlimited i.e. the spectral contents are confined to a maximum frequency $[\omega_{max}]$. Bandlimited signals can be perfectly reconstructed from equispaced samples with a rate at least twice the maximum frequency in the signal. This mathematical framework known as Whittaker- Nyquist-Kotelnikov-Shannon (WNKS) Theorem. The  theorem sets forth the number of measurements required to reconstruct a band limited signal. Signal acquisition and processing has witnessed transition from the analog to the digital domain , 'ridden the wave of Moore's law'  and has driven the development of digital data acquisition devices for more than half century on the basis of the theorem.

According to a recent study by International Data Corporation (IDC), Massachusetts  , the premier global provider of market intelligence, advisory services, and events for the information technology, telecommunications and consumer technology markets , our  digital universe (which is dominated by sensor data)  will grow to an almost inconceivable 35 Trillion Gigabytes or 35 Zeta Bytes by the year 2020 [1]. Moreover the expanding gap between sensor data production and available data storage means that 'sensor systems will increasingly face a deluge of data that will be unavailable later for further analysis'[2]. Also , exponentially expanding gaps exist between sensor data production and both computational power and communication rates. Efficient management,  and navigation  of this data deluge calls for , 'fundamental advances in the theory and practice of sensor design; signal processing algorithms; wideband communication systems; and compression, triage, and storage techniques'. 'In response to the resulting challenge , … signal-processing researchers have spent the last several decades creating powerful new theory and technology for digital data acquisition (digital cameras, medical scanners), digital signal processing (machine vision; speech, audio, image, and video compression), and data communication (high-speed modems , Wi-Fi) that have both enabled and accelerated the information age' [2].

The  well known scheme for digital data compression   is known as Transform Coding (TC) or  Lossy  Compression (LC) .The key factor which enables LC i.e. the mapping of the high-dimensional raw sensor data   to an extremely low-dimensional subset   is the observation that many natural signals are sparse or compressible in the sense that they have 'concise representations' or structures when expressed in  a transformed domain or basis . Transforms by themselves do not provide any compression . However , by reallocation of the energy in the data , transforms provide the possibilities for compression. 'Adaptive quantization and entropy coding when applied to these transform coefficients results in significant reduction in bit rates facilitating data transmission in embedded bit stream format'.  Data Compression , therefore, is an essential component in today's world of massive data storage and transmission. The  well known methods for data compression  are  Discrete Cosine Transform  (DCT) , Discrete Sine Transform (DST) and Discrete Wavelet Transform (DWT) which provide sparse or compressible representations for signals and images in a class of interest.

By a *sparse representation*, we mean that we can represent a signal of length $n$ with $k \ll n$ nonzero coefficients; by a *compressible representation*, we mean that the signal is well approximated by a signal with only the $k$ nonzero coefficients. Both sparse and compressible signals can be represented with high accuracy by preserving only the values and locations of the largest $k$ coefficients of the signal. This approach which exploits both sparsity and compressibility is also known as sparse approximation and forms the foundations for standards viz. jpeg, jpeg2000, mpeg , used for compression of images and audio signals in the industry .

Mathematically , a signal $x$ is compressible if its sorted coefficient magnitudes $|\theta_n|$ in transform domain $\Psi$ follow the power law decay i.e. $|\theta_n| \leq n^{-q}$, $n = 1,2,\ldots$. The larger q is, the faster the magnitudes decay, and the more compressible a signal is. Fig 1 illustrates compressibility of a sample speech signal that follows $|\theta_n| \approx 46.97\, n^{-1.45}$, $n = 1,2,\ldots 2048$, in the Discrete Cosine Transform Basis. It can be easily verified that the Mean Square Error between the true signal and its $k \ll 2048$ term approximations [ i.e. by preserving only the $k$ coefficients with largest magnitudes , also known as *Hard Thresholding* ] , is of the order of $O(10^{-4})$, confirming that the signal is sparse and compressible. We define such a signal as *k-compressible*.

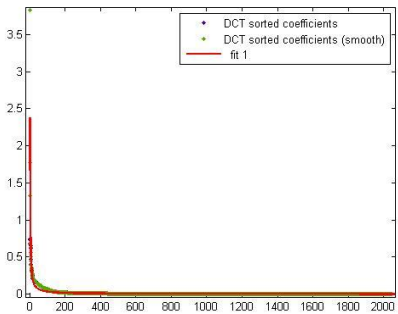

FIG 1. POWER LAW DECAY OF SORTED DCT COEFFICIENTS [ θ ] FOR A SAMPLE SEGMENT OF SPEECH FILE female1.wav ITU-T P.501(2009) © ITU.

An acquisition protocol , on the other hand , 'which performs *as if* it were possible to directly acquire just the important information about the signals/images—in effect, not acquiring that part of the data that would eventually just be "thrown away" by lossy compression'[9] has recently been proposed by David L. Donoho , Emmanuel J. Candés , Justin Romberg and Terrence Tao [3][4][5]. The neologism 'Compressed Sensing' (CS) was coined by Donoho in 2006 [3] to represent this sensing paradigm. Donoho , Candés , Romberg and Tao had mathematically established that 'a finite-dimensional signal having a sparse or compressible representation can be recovered from a small set of linear, non-adaptive incoherent measurements'[10]. The design of these measurement schemes and their extensions to practical data models and acquisition schemes are one of the most central challenges in the field of Compressed Sensing.

'CS differs from classical sampling in two important respects. First, rather than sampling the signal at specific points in time or space , CS systems typically acquire measurements in the form of inner products between the signal and a more general test function called Sensing Matrix. Randomness often plays a key role in the design of these test functions. Second, the two frameworks differ in the manner in which they deal with signal recovery, i.e., the problem of recovering the original signal from the compressive measurements'[10]. In the WNKS framework, signal recovery is achieved through 'cardinal ($Sinx/x$) interpolation - a linear process that requires little computation and has a simple interpretation' [10].

In this paper we have taken a segment of a speech file sampled at Nyquist rate and consider the inverse process viz. sparse recovery of the signal from non-adaptive measurements.

Following a brief introduction to the notions of sparse and compressible representation , the theory of compressed sensing is outlined in Section II. Numerical simulation of sparse recovery of a segment of the speech file *female1.wav* ITU-T P.501(2009) using DCT as the transform basis , an i.i.d Gaussian Random Matrix as the sensing matrix and $\ell_1$ norm minimization as the recovery algorithm , is described in section III. In the concluding section we briefly discuss some alternative transforms and algorithms . We also provide pointers for future research in compressed sensing.

Readers may also refer to the excellent tutorials , reviews and the references cited therein [8][9][10][11] for exploring this emerging sensing paradigm.

## II. THEORY OF COMPRESSED SENSING

Following definitions and theorem outline the mathematical foundations of compressed sensing which explore methodologies from various other fields viz. applied harmonic analysis, frame theory, functional analysis, numerical linear algebra, optimization theory, random matrix theory and probability.

**Definition 1. Sparsity [10]**

Let $x = \{x_i\}_{i=1}^{n} \in \mathbb{R}^n$ be the signal of interest , $\theta = \{\theta_i\}_{i=1}^{n} \in \mathbb{R}^n$ and $x = \Psi\theta$ in the basis or frame $\Psi$ , then $x$ is *k-sparse* iff ,

$$\|\theta\|_0 = \#\{k: \theta_k \neq 0\}, \; k \ll n \qquad (1)$$

where ,
  [ $\Psi = \Psi^{n \times n}$ Transform Matrix , and
    $\theta = \theta^{n \times 1}$ Coefficient Vector ]

Obviously , we will be dealing with signals that are not themselves sparse, but which admit a sparse representation in some basis $\Psi$. 'Sparsity figures prominently in the theory of statistical estimation and model selection, in the study of the human visual system, and has been exploited heavily in image processing tasks, since the multi-scale wavelet transform provides nearly sparse representations for natural images' [10].

**Definition 2. Nonadaptive Linear Measurements [10]**

Given a signal $x = \{x_i\}_{i=1}^n \in \mathbb{R}^n$ defined by (1), $m$ nonadaptive linear measurements $y = \{y_i\}_{i=1}^m \in \mathbb{R}^m$ are defined by :

$$y = \Phi x \;;\; [\Phi = \Phi^{m \times n}, \text{ is the Sensing Matrix}] \quad (2)$$

The sensing matrix $\Phi$ represents dimensionality reduction because it maps $\mathbb{R}^n$ into $\mathbb{R}^m$, where $m \ll n$. In the framework of CS it is assumed, in general, that the measurements are non-adaptive, meaning that the rows of $\Phi$ are fixed in advance and do not depend on the previously acquired measurements.

**Definition 3. Mutual Coherence [6]**

The coherence between the sensing basis $\Phi$ and the representation basis $\Psi$ is defined as :

$$\mu(\Phi, \Psi) = \sqrt{n} \max_{i \leq k, j \leq n} |<\varphi_k, \psi_j>| \quad (3)$$

and measures the largest correlation between any two elements of sensing matrix $\Phi$ and $\Psi$. $\mu(\Phi, \Psi)$ is bounded by $[1, \sqrt{n}]$, when the columns of both $\Phi$ and $\Psi$ are normalized. Compressed sensing is concerned with low coherence pairs. Random matrices are largely incoherent with any fixed representation basis $\Psi$ such as DCT and DWT. If $\Phi$ is an orthobasis selected uniformly at random, then the coherence between $\Phi$ and $\Psi \approx (2 \log n)^{1/2}$ [7]. Thus, Best choice for $\Phi$ are random matrices such as i.i.d Gaussian and Bernoulli Matrix.

The CS recovery process consists of a search for the sparsest signal $x$ that yields the measurements $y$.

Defining $\ell_0$ norm of a vector $\|x\|_0$ as the number of nonzero entries in $x$, the simplest way of posing a sparse recovery algorithm, in this context, is by using the following optimization scheme

$$\min_\theta \|\theta\|_0 \quad \text{subject to} \quad y = \Phi\Psi\theta \quad (4)$$

Since $x$ is $k$ sparse, θ must belong to one of $^nC_k$ subspaces in $\mathbb{R}^n$. Similarly, $y$ must belong to one of $^mC_k$ subspaces in $\mathbb{R}^m$. For almost all $\Phi^{m \times n}$ with $m \geq k + 1$, an exhaustive search through the subspaces can determine which subspace $x$ belongs to and thereby recover the signal's sparsity pattern and values. Therefore, in principle, a $k$ sparse signal can be recovered from as few as $m = k + 1$ random samples [11]. However the exhaustive search is NP Complete i.e. computationally intractable for even moderately large values of $k$ and $n$. 'Our goal, therefore, is to find computationally feasible algorithms that can successfully recover a sparse vector from the measurement vector $y$ for the smallest possible number of measurements' [9].

**Theorem1. Random Incoherent Sampling Theorem [6]**

If $x$ is $k$-sparse in the basis or frame $\Psi$ and if $m$ measurements $y$ in the $\Phi$ domain are given uniformly at random, then if

$$m \geq c \, \mu^2(\Phi, \Psi) \, k \log n \quad (5)$$

then the signal $x$ within the class of interest such that $y = \Phi x$, could be recovered with overwhelming probability by solving the convex optimization problem i.e. $\ell_1$- norm minimization or Basis Pursuit [3][4][5],

$$\min_\theta \|\theta\|_1 \quad \text{subject to} \quad y = \Phi x = \Phi\Psi\theta \quad (6)$$

For incoherent bases $\mu^2(\Phi, \Psi) \approx 1$. Hence, $m \geq c \, k \log n$, which provides a lower bound on the number of incoherent measurements $m$. The scheme is numerically stable and robust against noise, while requiring a number of measurements $m$ comparable to $k$, the sparsity level.

The core of the CS problem is to determine when $\ell_0 = \ell_1$. Well known sufficient conditions for this to hold true are Mutual Coherence and Reduced Isometry Property [11].

Sparsity expresses the idea 'that the 'information rate' of a signal may be much smaller than suggested by its bandwidth' as explained in Section I , 'incoherence extends the duality between time and frequency and expresses the idea that the objects having a sparse representation in the representation basis $\Psi$ must be spread out in the domain in which they are acquired, just as a Dirac or spike in the time domain is spread out in frequency domain' [6].

Reduced Isometry Property [RIP] , initially introduced in [4], measures the degree to which each subset of $k$ column vectors of $\Phi$ is close to being an isometry and is defined in the followings.

**Definition 4. Reduced Isometry Property [4]**

The sensing matrix $\Phi^{m \times n}$ has RIP of order $k$, if there exists a

$\delta_k \in (0,1)$ such that

$$(1-\delta_k)\|x\|_2^2 \leq \|\Phi x\|_2^2 \leq (1+\delta_k)\|x\|_2^2 \quad \text{for all } x \in \Sigma_k \quad (7)$$

The set $\Sigma_k$ contains all signals $x$ that are $k$-sparse. When this property holds $\Phi$ approximately preserves the Euclidean length of k sparse signals, which in turn implies that k-sparse vectors cannot be in the null space of Φ.

An equivalent description of the RIP is to say that all subsets of k columns taken from Φ are in fact nearly orthogonal [7]. The condition that Φ must satisfy RIP is also necessary from the standpoint of Projective Geometry.

### III. SPARSE RECOVERY OF SPEECH SIGNALS FROM NONADAPTIVE GAUSSIAN MEASUREMENTS

A 5.3 ms segment of 'female1.wav ITU-T P.501(2009)', sampled at 48kHz was selected for our numerical simulation. Discrete Cosine Transform has been used as the transform basis Ψ. Sparsity level of the signal in the transformed domain was selected at $k = 128$ with reference to the power law decay of the transformed coefficients shown in Fig1. The efficiency of the transform in cumulatively reallocating the signal energy in the transformed domain is also important in this context. For sensing, an i.i.d Gaussian Random Matrix (0,0.02) was selected [ $\Phi = \Phi^{512 \times 2048}$ ]. In Fig 2 the original signal has been compared with the $\ell_1$-norm recovered signal.

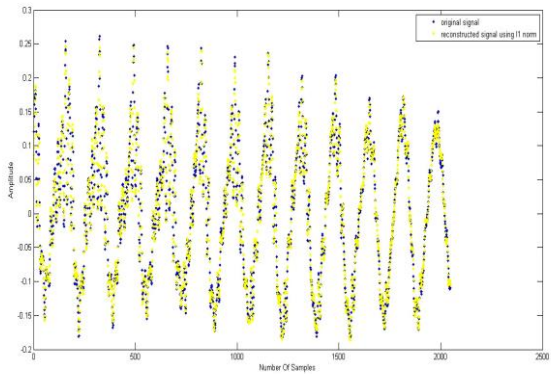

FIG 2. ORIGINAL SIGNAL(**BLUE**) VS $\ell_1$ RECOVERED SIGNAL(**YELLOW**) FOR A SEGMENT OF SPEECH FILE 'FEMALE1.WAV ITU-T P.501(2009)' © ITU.

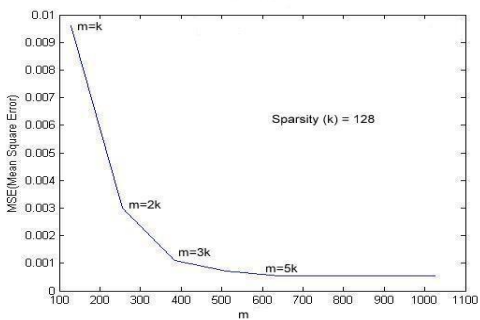

FIG 3. ASYMPTOTIC CONVERGENCE OF MSE FOR THE $\ell_1$ RECOVERED SIGNAL

Convergence characteristics of $\ell_1$ norm minimization algorithm is shown in Fig 3. In this figure MSE of the sparse recovered signal is plotted for values of $m$ which are integral multiples of $k$, the sparsity level. It may be observed that the MSE asymptotically converges for $m \geq 4k$ to $O(10^{-4})$, validating empirically the 'de facto a known four to one practical rule'[6] i.e. for exact recovery one needs about four incoherent measurements per unknown non zero term, as specified in Theorem 1.

Finally in Fig 4 we compare (i) MSE between the original and $k$ term approximate i.e. IDCT signal and (ii) MSE between the original and the $\ell_1$-norm minimized or sparse recovered signal for various compressions. These plots show that MSE between the original and the recovered signal using either (i) k-term approximation or (ii) compressed sensing are of the order of $10^{-4}$ and almost identical at all levels of compression, thereby ensuring stability of the CS framework for signal recovery from incoherent random measurements.

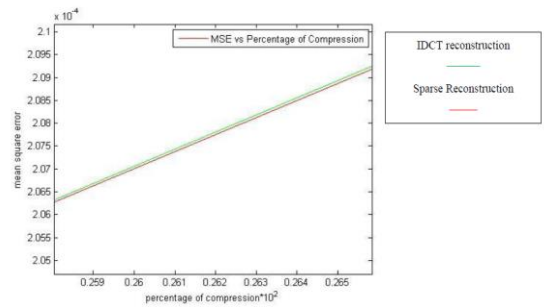

FIG 4. MSE BETWEEN THE ORIGINAL AND RECOVERED SIGNALS USING TRANSFORM CODING AND COMPRESSED SENSING

### IV. CONCLUSION

Though sparse recovery of a signal from nonadaptive linear measurements appears as a poorly defined problem and a computationally impractical goal in general, mathematical analysis [3][4][5][10] together with numerical simulations outlined in the paper establishes that subject to the conditions of *Incoherence* and *Reduced Isometry Property* we can obtain unique and stable solution via $\ell_1$ minimization.

CS initially applied in paediatric MRI while preserving diagnostic quality by Candés, Romberg and Tao [4], has already notable impact on several applications. Moreover, 'the broad applicability of this framework has inspired research that extends the CS framework by proposing practical implementations for numerous applications' [10], e.g. sub-Nyquist analog-to-digital converters (ADCs), Compressive imaging architectures, Compressive sensor networks, Linear regression and model selection, Sparse error correction, Group testing and data stream algorithms, Single-pixel camera, Hyperspectral imaging ,Compressive processing of manifold-modelled data, Inference using compressive measurements, and Genomic sensing.

It should be naturally enquired - whether there are alternative transforms, sensing matrices and algorithms that might also find the correct solution. 'It has been found that for certain media types e.g. musical sound with strong harmonic content , sinusoids are best for compression , noise removal and deblurring ; while for other media types e.g. images with strong edges , wavelets are a better choice than sinusoids' [8]. DWT has been extensively used in CS for sparse recovery of images [8].

'Initial work in CS has emphasized the use of randomized sensing matrices whose entries are obtained independently from a standard probability distribution, such matrices are often not feasible for real-world applications due to the cost of multiplying arbitrary matrices with signal vectors of high dimension' [9] . Also, neither Gaussian nor Bernoulli are structured , they are not applicable in large scale problems. Available alternatives for structured CS matrices has been briefly reviewed by Duarte and Eldar [9]. Candés , Romberg and Tao used Random Partial Fourier Matrix [4] in reconstructing Biomedical Image using $\ell_1$ norm minimization. Sub Gaussian Random Matrices have also been mathematically established to satisfy RIP [10].

From the algorithmic point of view in addition to Convex Relaxation Techniques e.g. Basis Pursuit , Greedy Methods e.g. Orthogonal Matching Pursuit and Combinatorial Techniques have been successfully applied for sparse recovery. Candes , Wakin , Boyd used reweighted $\ell_1$ minimization that 'in many situations outperforms $\ell_1$ norm minimization in the sense that substantially fewer measurements are required for exact recovery' [12]. Both sensing matrix design and recovery algorithms are active areas of research in the context of CS.

In conclusion we can say that 'the potential pay-offs of CS are huge', as 'removing the Nyquist barrier in the resolution limited applications' 'can improve the user experience, increase data transfer, improve imaging quality and reduce exposure time – in other words, make a prominent impact on the analog-digital world surrounding us' [9] and could ' enable radically new information technologies and powerful new tools for scientific discovery' [2], in near future.

ACKNOWLEDGMENT

The authors are grateful to Carlos A. Sing-Long , ICME, Stanford University , California for critically reviewing the manuscript and offering valuable comments on the contents of the paper .